\documentclass[preprint,12pt]{elsarticle}

\usepackage{amsmath}
\usepackage{color}
\usepackage{url}
\usepackage{subcaption}  
\usepackage{array} 
\newcolumntype{L}[1]{>{\raggedright\let\newline\\\arraybackslash\hspace{0pt}}m{#1}}
\newcolumntype{C}[1]{>{\centering\let\newline\\\arraybackslash\hspace{0pt}}m{#1}}
\newcolumntype{R}[1]{>{\raggedleft\let\newline\\\arraybackslash\hspace{0pt}}m{#1}}
\usepackage[normalem]{ulem} 

\definecolor{gray}{RGB}{90,90,90}
\newcommand{\removed}[1]{} 
\newcommand{\grayed}[1]{} 
\newcommand{\added}[1]{#1}

\newcommand{\comment}[1]{}

\usepackage{lineno}
\usepackage{hyperref} 

\journal{Nuclear Instruments and Methods A}

\begin{document}

\begin{frontmatter}



\title{GEANT4 models of HPGe detectors for radioassay}


\author[pnnl]{R.H.M.~Tsang  \corref{cor}}
\author[ua]{A.~Piepke}
\author[ua]{D.~J.~Auty \fnref{dja}}
\author[laurentian]{B.~Cleveland \fnref{bc}}
\author[bern]{S.~Delaquis \fnref{dead}}
\author[ua]{T.~Didberidze}
\author[ua]{R.~MacLellan \fnref{rm}}
\author[ua]{Y.~Meng \fnref{ym}}
\author[ua]{O.~Nusair}
\author[bern]{T.~Tolba \fnref{tt}}

\cortext[cor]{Corresponding author, heiman.tsang@pnnl.gov}
\fntext[dja]{Now at University of Alberta}
\fntext[bc]{Also at SNOLAB}
\fntext[dead]{Deceased}
\fntext[rm]{Now at University of South Dakota}
\fntext[ym]{Now at Shanghai Key Laboratory for Particle Physics and Cosmology, Institute of Nuclear and Particle Physics (INPAC) and School of Physics and Astronomy, Shanghai Jiao Tong University, Shanghai 200240, China}
\fntext[tt]{Now at Institut f\"ur Kernphysik, Forschungszentrum J\"ulich, 52428 J\"ulich, Germany}

\address[pnnl]{Pacific Northwest National Laboratory, Richland, WA 99352}
\address[ua]{Department of Physics and Astronomy, University of Alabama, Tuscaloosa, AL 35487}
\address[laurentian]{Department of Physics, Laurentian University, Sudbury, Ontario P3E 2C6 Canada}
\address[bern]{LHEP, Albert Einstein Center, University of Bern, Bern, Switzerland}

\begin{abstract}
Radiation transport models of two high purity germanium detectors, 
GeII and GeIII, located at the University of Alabama have been created in GEANT4 \cite{geant4}.
These detectors have been used extensively for radioassay measurements of materials used in various low background experiments.
The two models have been validated against actual data under several scenarios typically seen in radioassay measurements.
The systematic uncertainties of the models for GeII and GeIII are estimated to be $\sim$12\% and \added{$\sim$9\%} respectively.
\end{abstract}


\end{frontmatter}


\section{Introduction}
\label{sec:intro}

Rare event searches such as 
neutrinoless double beta decay and dark matter experiments 
require low background level to reach high detection sensitivity. 
One of the major sources of background is 
gamma ray emissions from long-lived natural radioactivity,
e.g. U and Th,
present in the detector construction materials.
In order to accurately estimate the background rate of the detector, 
radioassays of the construction materials are required.

While various techinques, including 
inductively coupled plasma mass spectrometry and 
glow discharge mass spectrometry,
are often employed for such an application, 
gamma spectrometry has several advantages that stands out among them. 
\begin{itemize}
\item Gamma spectrometry directly observes gamma rays emitted by the nuclides creating background,
the emission rate of which is used to infer the activity of the radionuclide in the sample. 
This requires no assumptions on isotopic abundance, as is typically required for techniques 
that measure elemental concentration of higher members of a decay sequence as a proxy.
Taking full advantage of this method does require large, typically kg-size, samples.
Eventually, self absorption of the sample limits the gain to be made by further enlarging the counted sample.
\item It is non-destructive, and hence suitable for radioassays of final parts 
to be assembled in the detector, or valuable materials that cannot be easily replaced or procured (e.g. ancient lead).
\item It can reach high sensitivity (in the order of $\mu$Bq/kg) when coupled with neutron activation. 
Neutron activation transmutes nuclides with long halflives to those with much shorter halflives, 
thus boosting their specific activities.
\end{itemize}

Because of variable size and shape of the large counting samples,
a challenge that faces gamma spectrometry is 
the accurate determination of detection efficiencies.
Detection efficiency is often defined as the 
``full absorption peak efficiency'' -- the probability of a gamma ray emitted by the sample
depositing all its energy in the germanium crystal,
taking full advantage of the excellent energy resolution of Ge detectors.
This depends on the attenuation length and the geometrical acceptance of the germanium crystal and 
attenuation of the gamma rays over its propagation through the sample.
Analytical calculation is only feasible in the simplest cases.
Typically, radiation transport simulations are required to determine detection efficiencies.

This paper describes 
the GEANT4 models of
two high purity germanium (HPGe) detectors located at the University of Alabama (UA),
and their validations.

\section{Description of the Ge detectors}
\label{sec:det}

\begin{figure*}
\centering
\includegraphics[width=\textwidth]{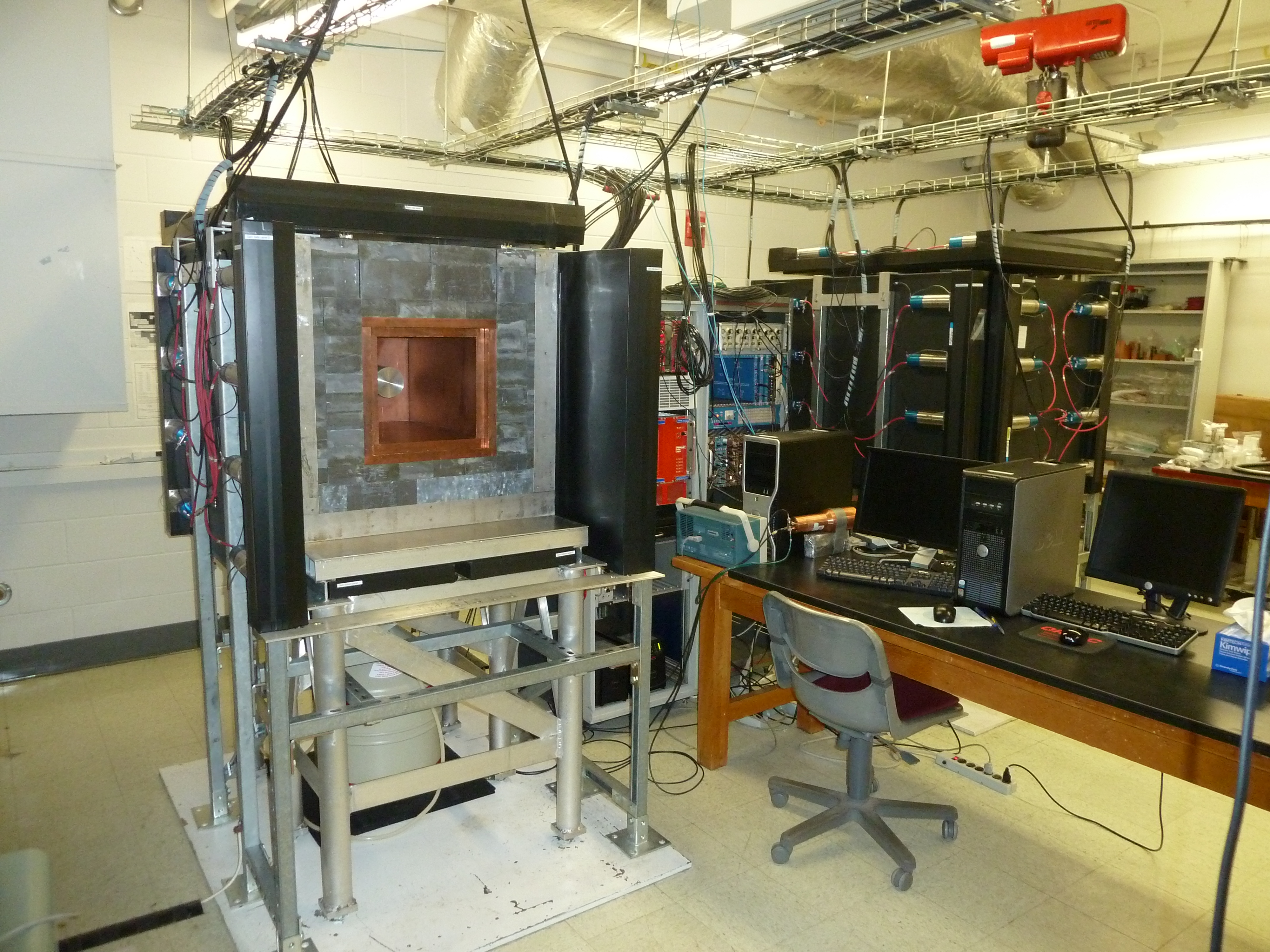}
\caption{
\added{
Photo of the UA counting lab: 
GeIII is shown on the left with its front shielding and plastic scintillation muon veto system removed for sample insertion;
GeII, with its shielding and muon veto closed, can be seen on the right.
}
}
\label{fig:ualabphoto}
\end{figure*}

The two HPGe detectors, dubbed GeII and GeIII,
are located on the ground level 
of Gallalee Hall 
at the University of Alabama, as pictured in Figure
\ref{fig:ualabphoto}.

The dimensions of the two detectors are shown in Table \ref{tab:gedims}.

\subsection{Ge detectors}
GeII is a Canberra GC6020 p-type coaxial germanium detector. 
The volume of the Ge crystal is about 261 cm$^3$.
The nominal thickness of the dead layer is 900 $\mu$m.
The Ge crystal is enclosed in an endcap with a diameter of 89 mm, made of oxygen-free copper.
GeII has been running since 2001.

GeIII is a Canberra GC10023 p-type coaxial germanium detector.
The volume of the Ge crystal is about 407 cm$^3$.
The nominal thickness of the dead layer is 700 $\mu$m.
The Ge crystal is enclosed in an endcap with a diameter of 95.25 mm, made of low activity aluminum.
This provides a higher sensitivity to gamma rays below 100 keV than GeII.
GeIII has been running since 2011.

Each Ge detector is surrounded by two layers of 1" (25.4 mm) thick copper plates, 
forming a 20" $\times$ 12" $\times$ 12" (508 $\times$ 304.8 $\times$ 304.8 mm) counting chamber.
Each of the two copper plates on the sample-insertion side of the chamber have four screw holes
for attaching two handles. 
This provides access to the chamber for sample placement and retrieval.
The copper inner shielding is surrounded by an 8" (203.2 mm) thick layer of lead bricks
\added{ to act as an outer shielding against ambient gamma radiation}.

\begin{table}
\centering
\begin{tabular}{|l|c|c|}
\hline
 & GeII & GeIII \\
\hline
\multicolumn{3}{|c|}{Ge Crystal} \\
\hline
Diameter & 70.5 & 80 \\
Length & 68 & 82 \\
Dead layer thickness (nominal) & 0.9 & 0.7 \\
Dead layer thickness (adjusted) & 1.42 & 0.7 \\   
\hline
\multicolumn{3}{|c|}{Crystal Holder} \\ 
\hline
Material & Cu & Cu \\
Thickness & 1.0 & 0.8 \\
\hline
\multicolumn{3}{|c|}{End Cap} \\
\hline
Material & Cu & Al \\
Diameter & 89 & 95.25 \\
Length & 140 & 159 \\
Entrance thickness & 1.0 & 1.5 \\
Side thickness & 1.5 & 1.5 \\
Ge front to endcap distance & 4.5 & 5.5 \\
\hline
\multicolumn{3}{|c|}{Shielding} \\
\hline
Inner Cu thickness & 50.8 \added{(2")} & 50.8 \added{(2")}\\
Outer Pb thickness & 203.2 \added{(8")} & 203.2 \added{(8")}\\
\hline
\multicolumn{3}{|c|}{Performance} \\
\hline
Relative Efficiency at 1.33 MeV & 60\% & 100\% \\
\hline
\end{tabular}
\caption{Some selected \added{parameters} \removed{dimensions} of GeII and GeIII. 
All \added{parameters} \removed{dimensions} come from \removed{manufacturer's} \added{manufacturers'} specifications \added{without associated uncertainties}, 
except for the dead layer thickness for GeII, 
which has been adjusted as described in Section \ref{subsec:tune}.
\added{The shielding thicknesses were originally specified in inches.}
All dimensions are in mm \added{unless specified otherwise}.}
\label{tab:gedims}
\end{table}

\subsection*{2.2. Calibration}
\label{sec:effcal}

The energy scale, resolution, and detection efficiency are determined by calibration runs with button sources placed 
at predefined locations in the counting chamber. 
Calibrations are performed on a regular basis.

First, the full absorption peaks, corresponding to the gamma energies $E_i$,
are identified in the spectrum,
then the peak for each gamma energy is fitted with the sum of a Gaussian 
and a linear background:
\begin{equation}
\label{eq:uncal}
f_{\mathrm{uncal.}}(C) = q_{i0} \cdot e^{-\frac{(C-\mu_i)^2}{2 \cdot s_{i}^2}} + q_{i1} \cdot C + q_{i2}
\end{equation}
where 
$C$ is the ADC channel variable, 
$\mu_i$ is the peak centroid, 
$s_i$ is the peak width, 
and $q_{i0}$, $q_{i1}$, and $q_{i2}$ are nuisance parameters.

\begin{figure}
\centering
\includegraphics[width=.8\textwidth]{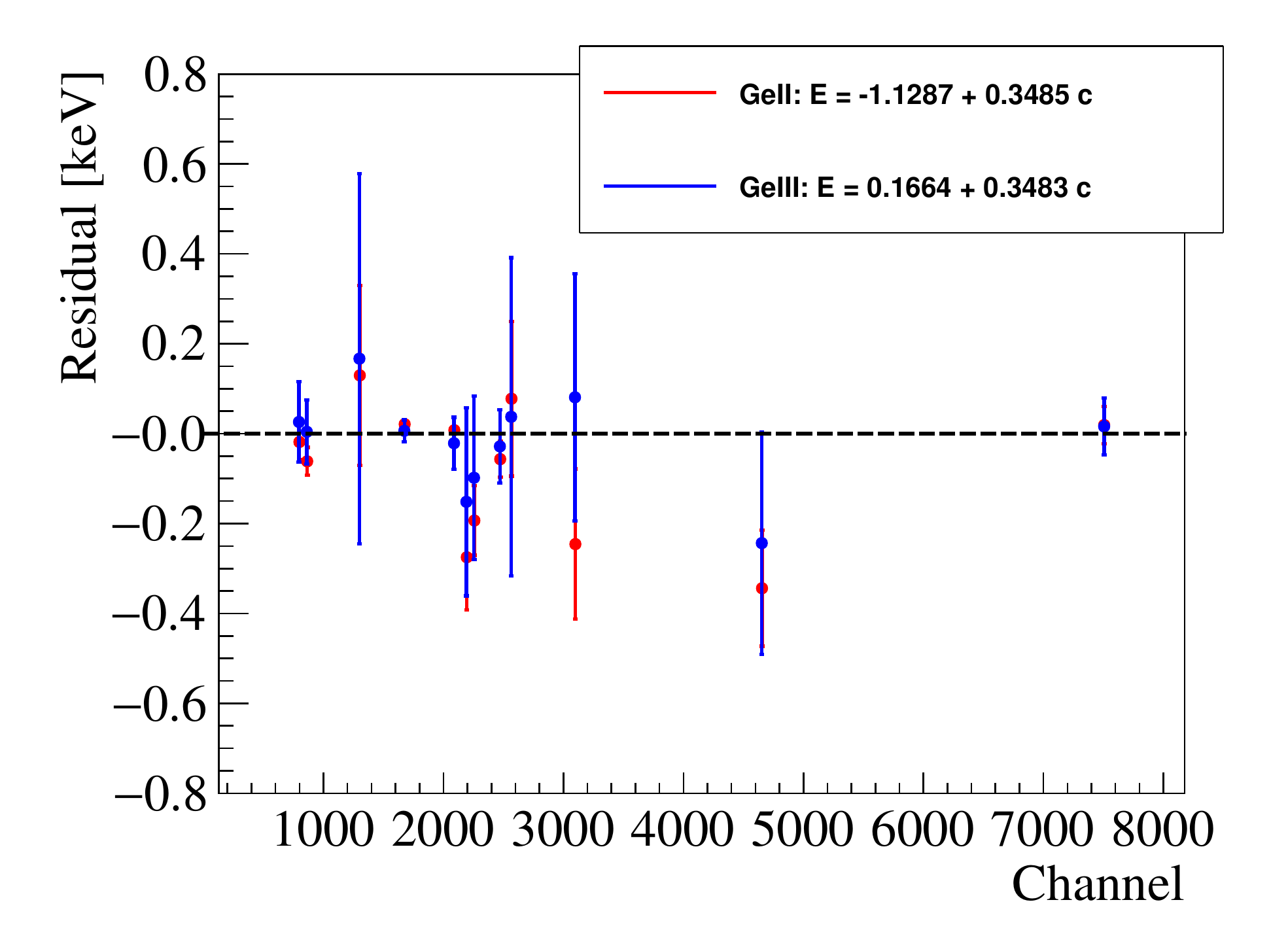}
\caption{Fit residuals (defined as $E_i - E(\mu_i)$) of a typical energy scale fit for 
GeII \added{($\chi^2=31.6$, ndf=10) in red} 
and 
GeIII \added{($\chi^2=2.5$, ndf=10) in blue}.
\added{Note that the error bars for some data points are too small to be seen. 
Low-event rate $^{228}$Th source data are shown. (Colors online)}
\comment{Now for both fits, only gamma lines from Th-228 are included. 
GeII used to contain lines from Ba-133, Co-60, Cs-137 and Th-228. 
GeIII used to contain lines from Ba-133 and Th-228.}
}
\label{fig:ge2enecal}
\end{figure}

The energy scale is found to be linear, and
is determined by fitting the following equation to the ($E_i$, $\mu_i$) pairs:
\begin{equation}
E(\mu) = m_\mu \cdot \mu + c_\mu.
\end{equation}
where $m_\mu$ and $c_\mu$ are free parameters.
An example is shown in Figure \ref{fig:ge2enecal}.

The peaks in the calibrated energy spectrum are then fitted to a function similar to Equation \ref{eq:uncal}:
\begin{equation}
f(E) = 
\frac{a_i}{m_\mu \cdot \sigma_i \cdot \sqrt{2\pi}} \cdot e^{-\frac{(E-E_i)^2}{2\cdot\sigma_i^2}} + b_i \cdot E + c_i
\end{equation}
where $a_i$ is the peak area,
$\sigma_i$ is the energy resolution, and $b_i$ and $c_i$ are the parameters for the linear background.
Using the fitted $a_i$, the detection efficiency is calculated as
$\varepsilon_i = \frac{a_i}{A_i \cdot r_i \cdot t}$ where 
$A_i$ is the activity of the source,
$t$ is the counting time,
and $r_i$ is the branching fraction of the gamma peak.

\begin{figure}
\centering
\includegraphics[width=.8\textwidth]{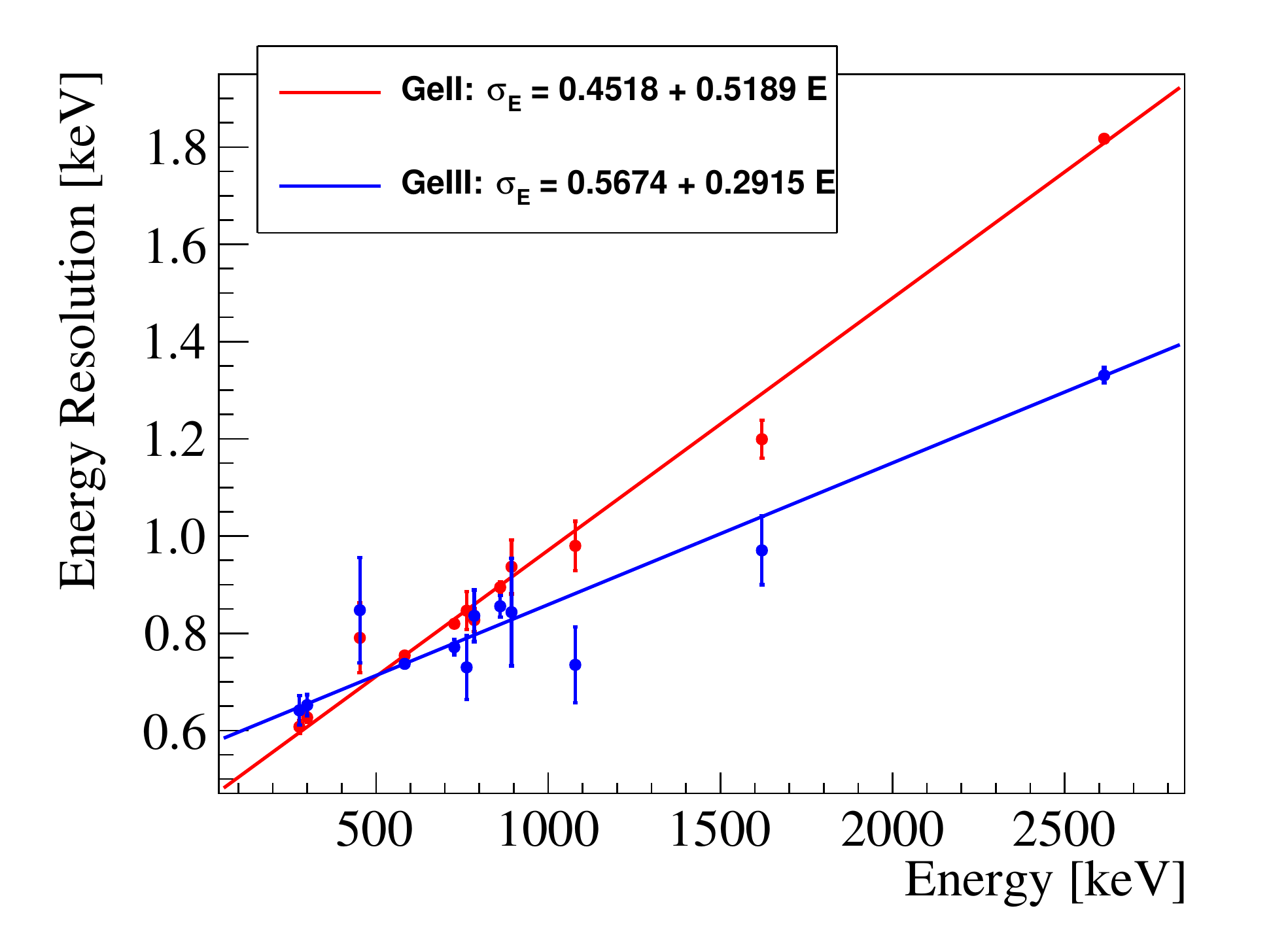}
\caption{A typical energy resolution fit for 
GeII \added{($\chi^2=17.1$, ndf=10) in red}
and 
GeIII \added{($\chi^2=10.9$, ndf=10) in blue}.
\added{Note that the error bars for some data points are too small to be seen.
Low-event rate $^{228}$Th source data are shown. (Colors online)}
\comment{Now for both fits, only gamma lines from Th-228 are included. 
GeII used to contain lines from Ba-133, Co-60, Cs-137 and Th-228. 
GeIII used to contain lines from Ba-133 and Th-228.}
}
\label{fig:ge2rescal}
\end{figure}

The resolution, as a function of energy, is also found to be linear, and
is determined by fitting the following equation to the ($E_i$, $\sigma_i$) pairs:
\begin{equation}
\sigma(E) = m_\sigma \cdot  E + c_\sigma.
\end{equation}
where $m_\sigma$ and $c_\sigma$ are free parameters.
An example of the fit is shown in Figure \ref{fig:ge2rescal}.

The detection efficiency is parameterized by the empirical relation below:
\begin{equation}
\label{eq:uaeff}
\varepsilon(E) = \frac{1}{E} \cdot \left[ p_0 + p_1 \cdot \ln(E) + p_2 \cdot \ln^2(E) + 
                             p_3 \cdot \ln^3(E) + p_4 \cdot \ln^5(E) + p_5 \cdot \ln^7(E)\right]
\end{equation}
where $p_0, p_1, \ldots ,p_5$ are free-varying parameters. 

\begin{table}
\centering
\begin{tabular}{|l|c|c|}
\hline
& GeII & GeIII \\
\hline
$m_\mu$ & 0.3485 & 0.3483 \\
$c_\mu$ & -1.128 keV & 0.1987 keV \\
$m_\sigma$ & 0.5598 & 0.2953 \\
$c_\sigma$ & 0.4544 keV & 0.5148 keV \\
\hline
\end{tabular}
\caption{Calibration constants used in the simulation code.}
\label{fig:calconsts}
\end{table}



\section{Implementation in GEANT4}
\label{sec:sim}

\subsection{Geometry}
All internal parts of GeII and GeIII and their endcaps are implemented in GEANT4 with boolean solids. 
The coded geometries are based on detailed technical information provided by the detector manufacturer for this purpose.
Both the copper and the lead shieldings are implemented. The muon veto system, however, is not included
as the scope of this work was, thus far, limited to the determination of counting efficiencies. 
A quantitative description of the detector backgrounds would require the modeling of the muon veto system.
Figure \ref{fig:ge23rendering} shows the GEANT4 renderings of GeII and GeIII.

\subsection{Readout}
For each simulated nuclear decay, all energy deposits in the active part of the germanium crystal are summed. 
To simulate the energy resolution of the actual detectors, the total energy is 
folded with a Gaussian resolution function with the following energy dependence:
\begin{equation}
E'= E + \eta \cdot \sigma(E)
\end{equation}
where 
$E$ is the total energy, 
$E'$ is the smeared total energy,
$\eta$ is a standard Gaussian random variable, and 
$\sigma(E)$ is the energy resolution of the actual detectors at $E$, 
as determined by calibration with radioactive sources.

The smeared total energy is mapped to the corresponding ADC channel ($C$) using the 
inverse of the energy calibration, mentioned above.
\added{Table 2 shows the parameters used in the simulation model for this purpose.}
This produces an uncalibrated energy spectrum similar to that delivered by the detector. 
This way, 
the same calibration and analysis procedure for actual data,
as described in Section \ref{sec:effcal},
can be applied to 
the simulated spectrum with minimal modification.

\subsection{Coincident gammas}
Coincident gammas are properly simulated by GEANT4 if the primary objects are 
the decaying nuclides as opposed to photons. 
In contrast, accidental coincidence of gamma rays originating from different nuclides are not considered by the simulation.

In practice, both kinds of coincidence have little effect in determining the detection efficiency.

\begin{figure}
\centering
\begin{subfigure}{\textwidth}
\centering
\includegraphics[clip, trim=4cm 4cm 0cm 4cm, width=.8\textwidth]{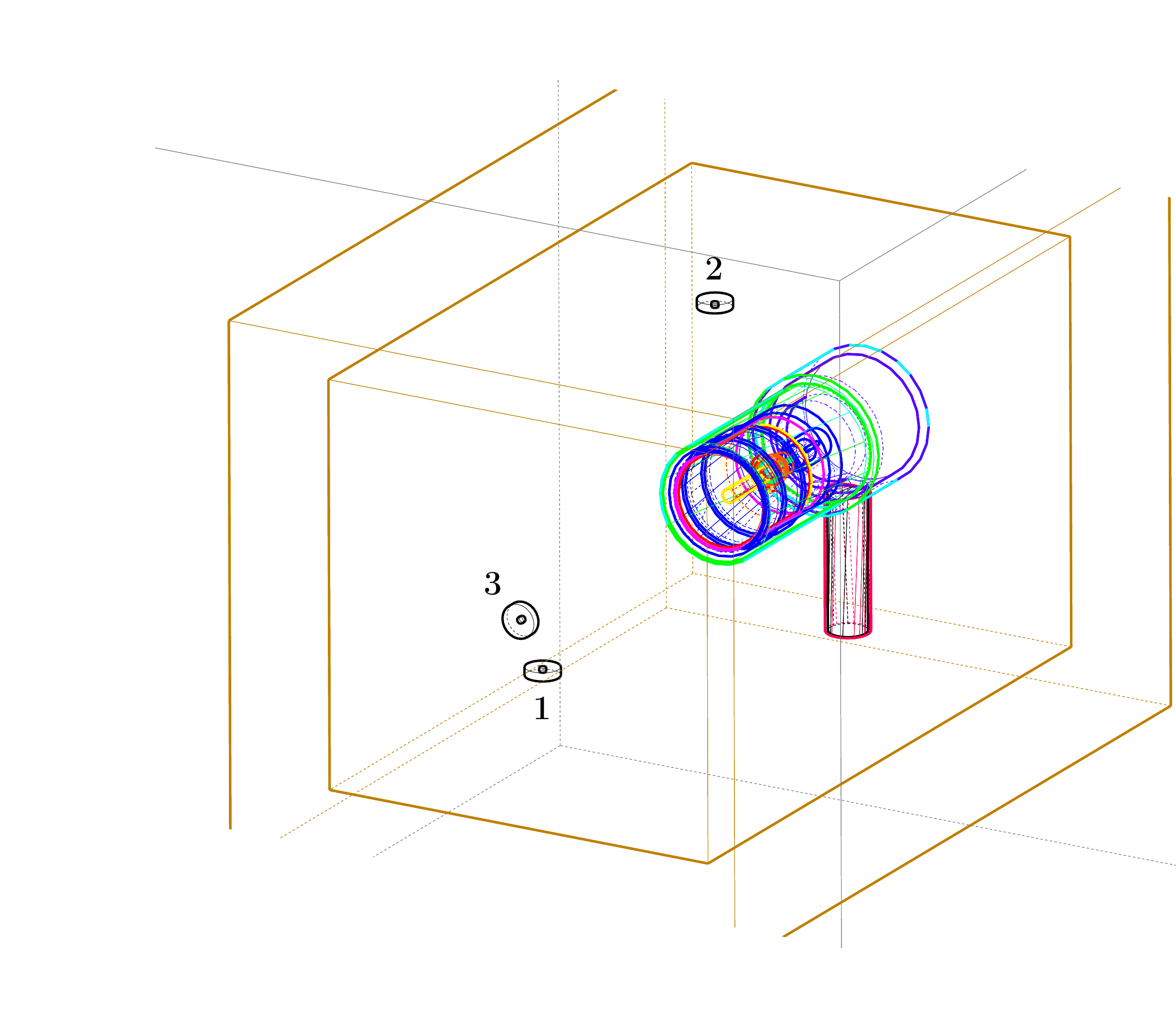} \\
\caption{GeII\\~\\}
\label{fig:ge2rendering}
\end{subfigure} %
\begin{subfigure}{\textwidth}
\centering
\includegraphics[clip, trim=4cm 4cm 0cm 4cm, width=.8\textwidth]{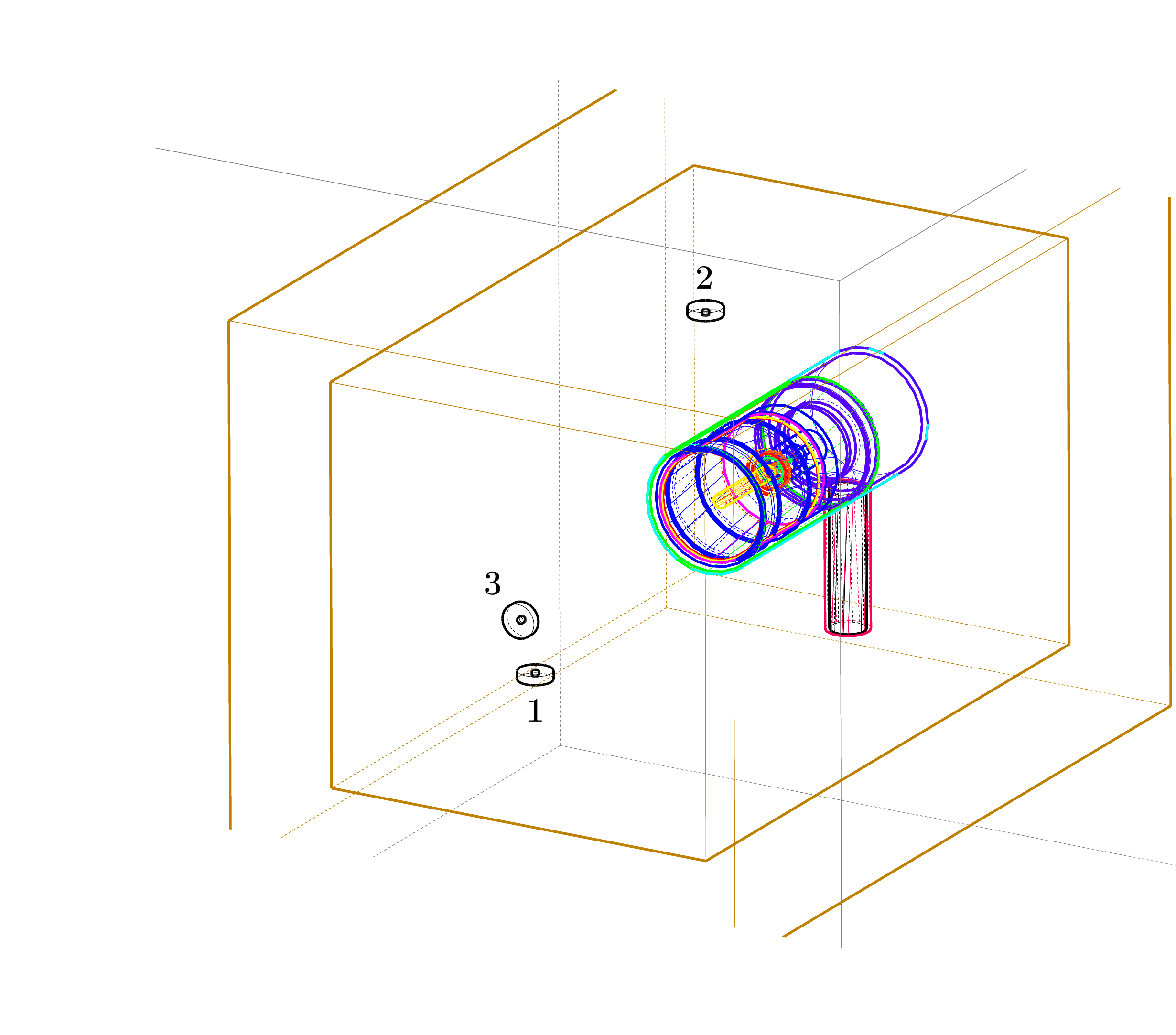} \\
\caption{GeIII}
\label{fig:ge3rendering}
\end{subfigure}
\caption{GEANT4 renderings of GeII (above) and GeIII (below). 
Also indicated on the renderings are the source positions labeled 1, 2 and 3. 
Positions 1 and 2 are flush with the detector front face. 
Position 1 is on the lower left edge of the chamber, 
while Position 2 is located under the ceiling of the chamber, above the central axis.
Position 3 is located right inside the chamber door, on the central axis of the detector.}
\label{fig:ge23rendering}
\end{figure}


\subsection{Dead layer tuning}
\label{subsec:tune}

Germanium crystals in p-type HPGe detectors typically have a ``dead layer'' on their outer surface
 that is insensitive to particle interactions. 
For GeII and GeIII, the nominal thicknesses of the dead layers, 
as specified by the manufacturer, are 0.9 mm and 0.7 mm, respectively. 
However, the thickness of the dead layer can change over time \cite{knoll}. 
Hence, an estimation of the dead layer thickness needs to be made 
in order to appropriately model it in the simulation,
especially at low energies (e.g. the modeling of $^{210}$Pb decays).

\begin{figure}
\centering
\begin{subfigure}{.8\textwidth}
\centering
\includegraphics[width=\textwidth]{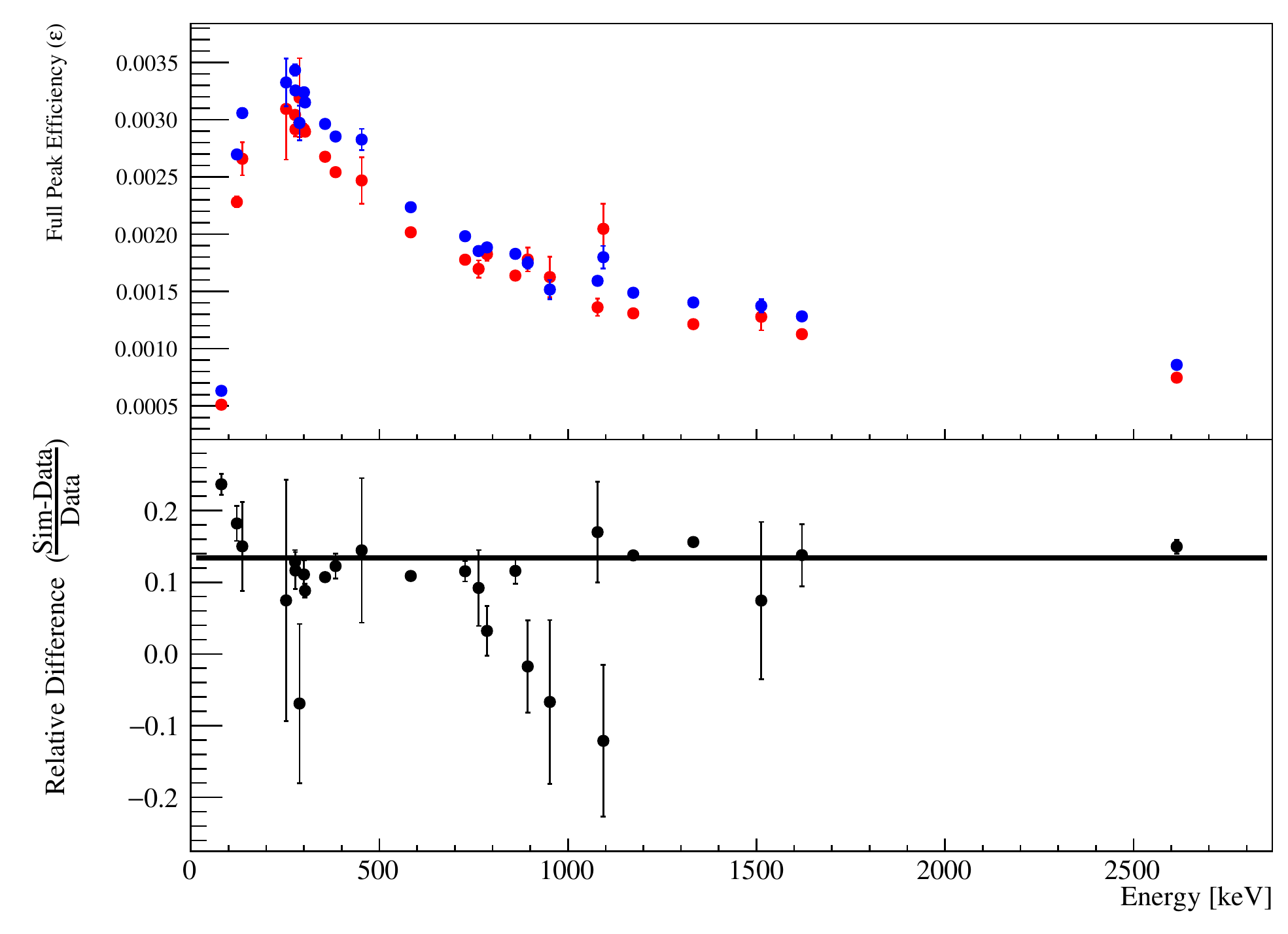}
\caption{Before dead layer tuning}
\label{fig:peaksagreement-ge2-a}
\end{subfigure}
\begin{subfigure}{.8\textwidth}
\centering
\includegraphics[width=\textwidth]{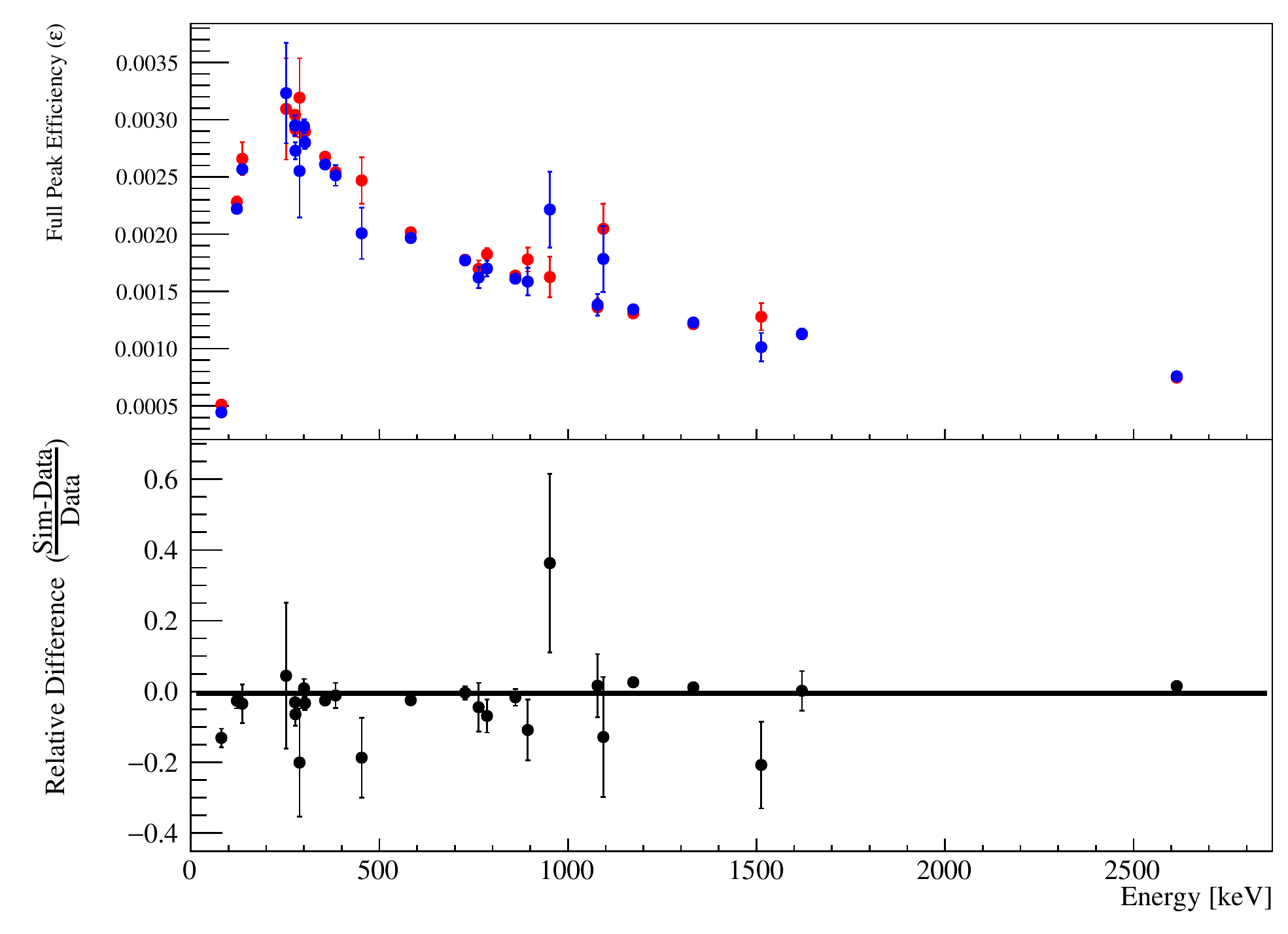}
\caption{After dead layer tuning}
\label{fig:peaksagreement-ge2-b}
\end{subfigure}
\caption{Relative difference in full absorption peak efficiency between simulation (blue) and data (red)
for GeII with the source placed at Position 1 as in Figure \ref{fig:ge2rendering}.
(Colors online)}
\label{fig:peaksagreement-ge2}
\end{figure}

\begin{figure}
\centering
\includegraphics[width=.8\textwidth]{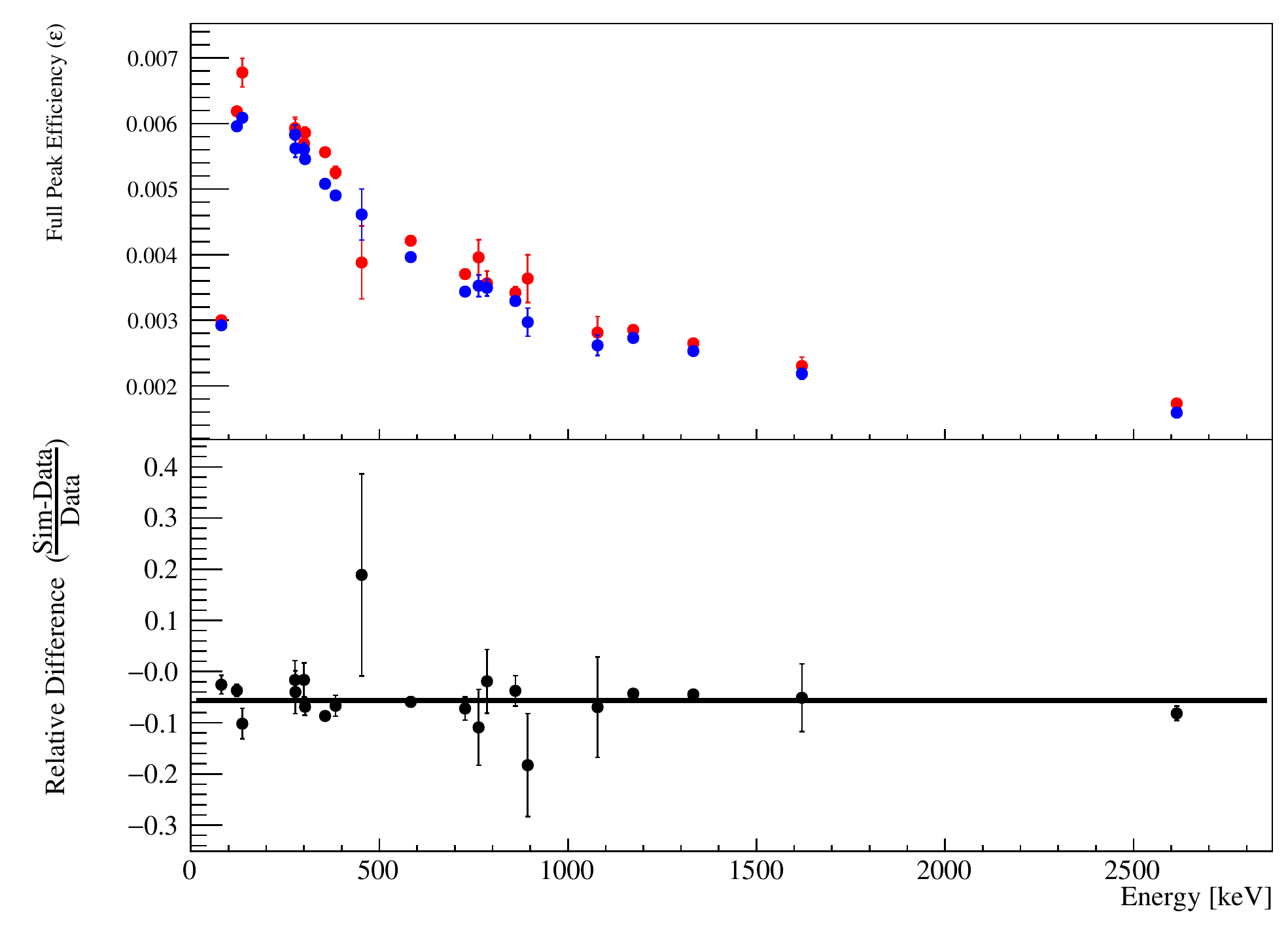}
\caption{Relative difference in full absorption peak efficiency between simulation (blue) and data (red)
for GeIII with the source placed at Position 1 as in Figure \ref{fig:ge3rendering}.
(Colors online)}
\label{fig:peaksagreement-ge3}
\end{figure}

To decide if the thickness of the dead layer needed to be tuned,
the detection efficiencies of GeII and GeIII were measured 
and compared with those calculated by the simulation assuming nominal dead layer thicknesses.
For the measurement, 4 button sources,
$^{228}$Th, $^{133}$Ba, $^{60}$Co, and $^{57}$Co, were placed in turn at Position 1 (as indicated in Figure 
\ref{fig:ge23rendering}). 
The results are shown in Figures \ref{fig:peaksagreement-ge2-a}
and \ref{fig:peaksagreement-ge3} for GeII and GeIII, respectively.

For GeIII, the simulation and the measurement show relatively small discrepancy ($<$10\%). 
However, for GeII, the calculated efficiency is systematically higher than the measured value.
There are two possible causes, both of which may contribute:
\begin{enumerate}
\item The actual dead layer thickness is larger than the nominal value. 
This causes additional attenuation to the gamma rays whose effect depends on the gamma energy. 
\item The geometry in the simulation does not accurately reflect the actual geometry.
In other words, the geometrical acceptance of the Ge crystal in the simulation differs from reality.
This effect is largely independent of the gamma energy.
\end{enumerate}

Cause (2) was indeed possible since 
the Ge detector and the cryostat, as one rigid body, can rotate around the penetration
through the copper and the lead shielding.
The Ge detector can therefore be accidentally nudged sideways by several millimeters
during sample placement or LN$_2$ filling. 
\added{The placement of the door can also affect the location of the source at position 3.}
\added{Therefore, even though}
\removed{Given that} the source positions are defined as fixed points in the chamber \removed{which is immobile},
it is possible that the actual geometry (which can change slightly from run to run) differs from 
that in simulation which is static.
This is sufficient to cause percent-level differences in detection efficiency.
However, this is not intrinsic to the detector, 
and can be corrected for by aligning the detector before sample placement or 
by attaching the sample onto the end cap.

To eliminate the effect of geometrical differences and to focus only on the dead layer thickness, 
a loss function ${L}$ \added{based on the mean squared error of the ratio of the measured and expected efficiencies} 
is defined as follows:
\begin{equation}
\begin{cases}
  \rho_i(t_+) =  \displaystyle\frac{\varepsilon_i}{\varepsilon(E_i) \cdot e^{-\mu(E_i) \cdot t_+}} \\
  {L}(t_+)  = \displaystyle\frac{1}{N_\gamma} \displaystyle\sum_{i=1} \rho_i^2(t_+) - 
              \left[\frac{1}{N_\gamma} \displaystyle\sum_{i=1} \rho_i(t_+)\right]^2
\end{cases}
\label{eq:chi2}
\end{equation}
where 
$t_+$ is the additional dead layer thickness beyond the nominal value,
$E_i$ is the energy of the $i$th gamma peak out of a total of $N_\gamma$,
$\varepsilon_i$ is the detection efficiency of the $i$th gamma peak as measured,
$\varepsilon(E_i)$ is the detection efficiency of the $i$th gamma peak calculated by simulation,
$\mu(E_i)$ denotes the attenuation coefficient for a photon with energy $E_i$ from the NIST XCOM database \cite{xcom}. 
The optimal $t_+$ is the one that minimizes ${L}(t_+)$. 
Notice that a difference in geometrical acceptance would manifest as an energy-independent pre-factor
in $\rho_i(t_+)$,
(and hence in $L(t_+)$) which has no effect on the value of the optimal $t_+$. 

\begin{figure}
\centering
\includegraphics[width=.8\textwidth]{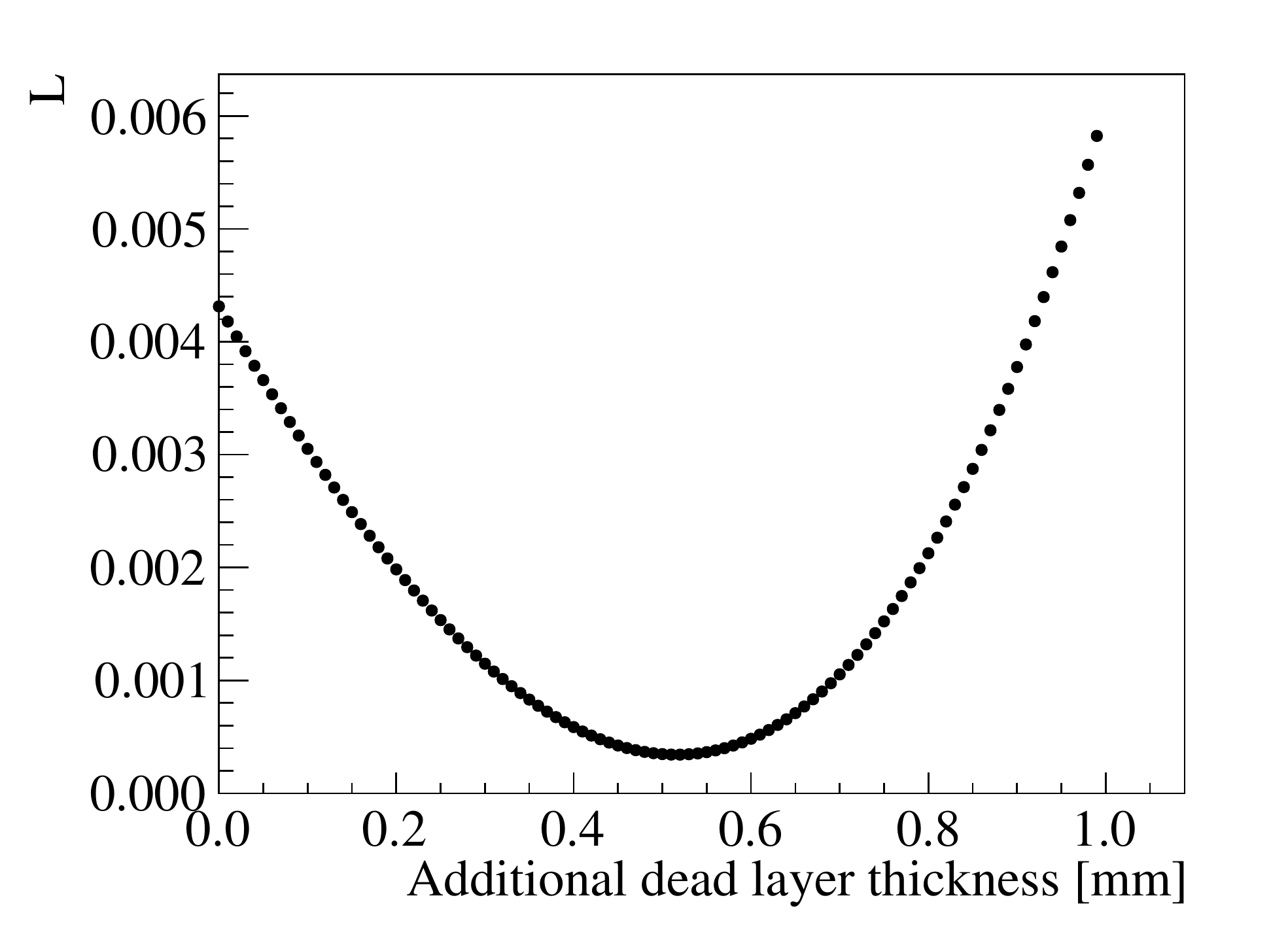} \\
\caption{$L$ as a function of additional dead layer thickness on top of the nominal value. 
It reaches a minimum at 0.52 mm.}
\label{fig:chi2}
\end{figure}

\begin{figure}
\centering
\includegraphics[width=.8\textwidth]{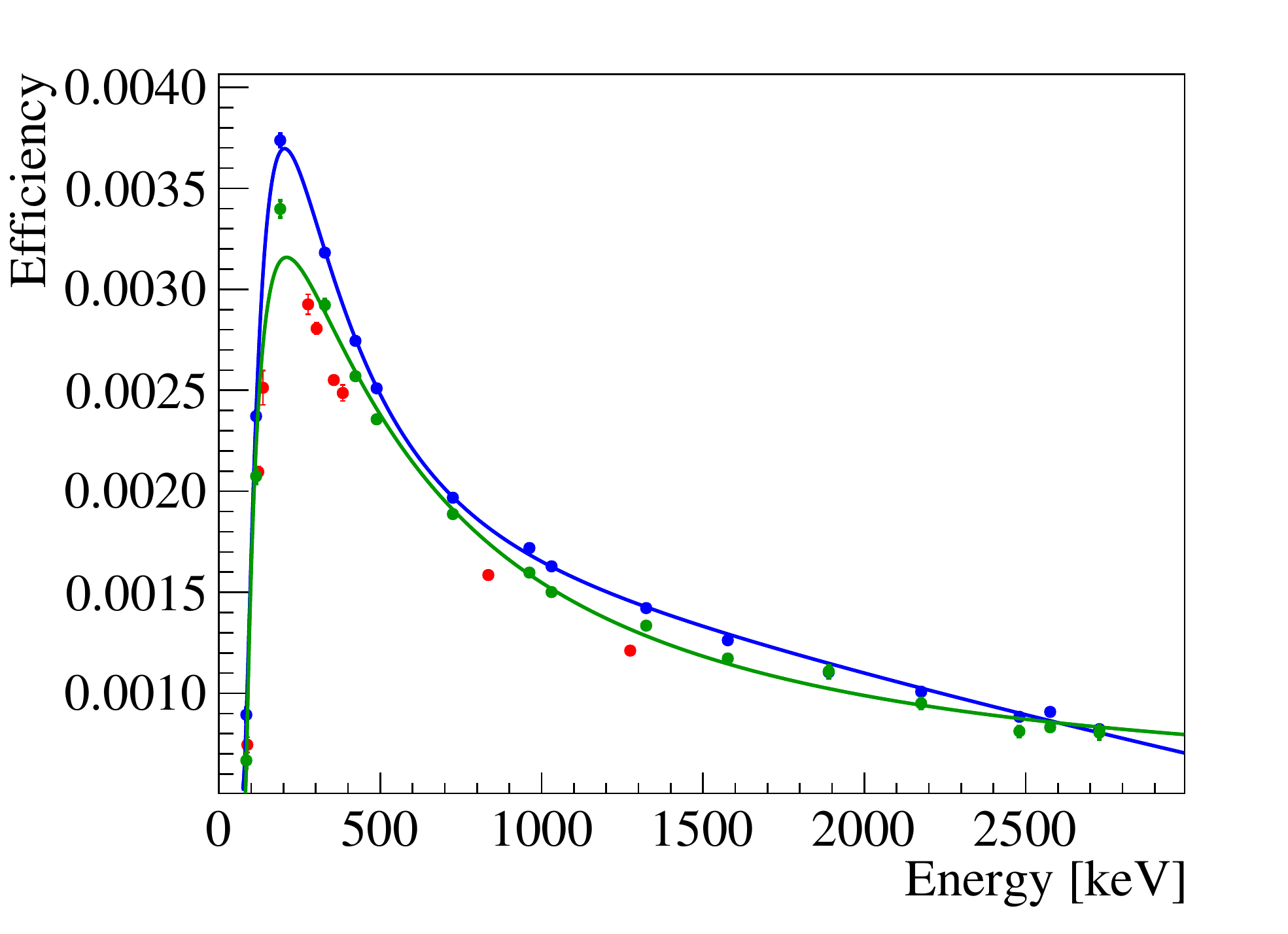} \\
\caption{This figure shows the effect of dead layer tuning for GeII. 
The data points in red are the detection efficiencies measured with 5 button sources. 
The data points in blue are the detection efficiencies calculated with the simulation model before dead layer tuning,
and the curve in blue is the fit. The data points and the curve in green are the same for after. \added{(Colors online)}}
\label{fig:tuning-eff}
\end{figure}

The detection efficiency of GeII was measured with 5 button sources:
$^{133}$Ba, $^{109}$Cd, $^{57}$Co, $^{54}$Mn, and $^{22}$Na placed in turn at Position 1.
The measured and the simulated detection efficiency values were put into Equation \ref{eq:chi2}.
Figure \ref{fig:chi2} shows ${L}(t_+)$ as $t_+$ is tuned in 10 $\mu$m steps up to 1 mm.
${L}(t_+)$ reaches a minimum when $t_+ = 0.52$ mm. 
The dead layer thickness in the GEANT4 model of GeII was therefore adjusted from 0.9 mm to 1.42 mm.
No adjustments were made to the dead layer of GeIII.
Figure \ref{fig:tuning-eff} shows the efficiency curves for GeII before and after dead layer tuning,
along with the efficiency values measured with the five button sources.

\begin{table}
\centering
\begin{tabular}{|l|c|c|c|c|}
\hline
& Button Source & Button Source & Button Source & Source cocktail\\
& Position 1 & Position 2 & Position 3 & in a bottle \\
\hline
GeII (before tuning) & $13.4 \pm \added{0.2}$ & $12.1 \pm \added{0.2}$ & $3.7 \pm \added{0.3}$ & - \\
GeII (after tuning) & $-0.5 \pm \added{0.3}$ & $3.0 \pm \added{0.3}$ & $-3.0 \pm \added{0.5}$ & $11.4 \pm \added{0.3}$ \\
GeIII & $-5.7 \pm \added{0.3}$ & $0.7 \pm \added{0.3}$ & $-3.4 \pm \added{0.4}$ & $8.6 \pm \added{0.2}$ \\
\hline
\end{tabular}
\caption{Percentage differences between measured and simulated detection efficiencies for the button sources and 
the source cocktail in GeII and GeIII, calculated as described in Section \ref{sec:posdep}. 
Note that Position 1 was used to tune GeII's dead layer thickness. }
\label{tab:syserr}
\end{table}
\section{Validation}
\label{sec:val}

To ensure the model is robust against a change in sample location, sample geometry, and gamma energy, 
four sets of validations have been performed:
\begin{itemize}
\item Button sources located at different positions. 
      This is to validate the performance of the model against a change in solid angle. 
\item A source cocktail in a bottle. 
      This is to validate that the model properly handles extended sources in addition to point sources.
\item Lead(II) nitrate [Pb(NO$_3$)$_2$] solution, 
      for low photon energies. 
\item Silica (SiO$_2$) spiked with $^{238}$U, $^{232}$Th and $^{40}$K. 
      This serves as an independent cross-check of the efficiency correction, compared to other germanium detectors operated outside UA.
\end{itemize} 

\subsection{Position dependence: Button sources at different positions}
\label{sec:posdep}
The $^{228}$Th, $^{133}$Ba, $^{60}$Co, and $^{57}$Co button sources,
the same ones used for dead layer check, were placed at locations 2 and 3, 
as indicated in Figure \ref{fig:ge23rendering}.
The efficiencies were calculated as described in Section \ref{sec:effcal} and 
compared with the simulation. 
The agreement between simulation and measurement was quantified by 
the average relative difference between simulation and measurement,
weighted by the inverse square of the uncertainty for each peak.
These averages, as seen in Table \ref{tab:syserr}, 
indicate that 
simulated and measured efficiencies agree 
to within $\sim$5\%.

\subsection{Extended source: Source cocktail in a bottle}

To check the performance of the model for a source with an extended shape,
a ``source cocktail'' (with its contents listed in Table \ref{tab:liqsrc}) was created as follows:
\begin{enumerate}
\item An empty, clean 125 mL polyethylene (PE) bottle was filled with 4M HCl to about half full.
\item
A total of $420.2 \pm 0.8$  $\mu$L of a calibrated Eckert \& Ziegler source solution,
which contained
$^{109}$Cd, $^{139}$Ce,  $^{57}$Co, $^{60}$Co, $^{137}$Cs, $^{113}$Sn and $^{88}$Y dissolved in HCl,
was transferred to the PE bottle.
The transfer was performed in 10 iterations with a 50 $\mu$L syringe. 
\item The PE bottle was topped off with 4M HCl to the intended fill level.
\end{enumerate}

\begin{table}
\centering
\begin{tabular}{|l|c|c|c|c|c|}
\hline
Nuclide & $E_\gamma$ & $\gamma$ branching & Half-life & $\gamma$ emissivity [$\gamma$/s]\\
        & [keV]      & fraction           & [d]       & on Nov 22, 2013 \\ 
\hline
$^{57}$Co  & 14.4   & 0.0916 & 271.74 & 2.161 \\
$^{109}$Cd & 88.0   & 0.0364 & 461.4 & 86.12 \\
$^{57}$Co  & 122.1  & 0.8560 & 271.8 & 20.19 \\
$^{57}$Co  & 136.47 & 0.1068 & 271.8 & 2.519 \\
$^{139}$Ce & 165.9  & 0.8000 & 137.6 & 4.029 \\
$^{113}$Sn & 255.1  & 0.0211 & 115.1 & 0.0739 \\
$^{113}$Sn & 391.7  & 0.6497 & 115.1 & 2.599 \\
$^{137}$Cs & 661.7  & 0.8510 & 10980 & 176.3 \\
$^{88}$Y   & 898.0  & 0.937  & 106.6 & 4.288 \\
$^{60}$Co  & 1173.2 & 0.9985 & 1925.28 & 260.3 \\
$^{60}$Co  & 1332.5 & 0.9998 & 1925.28 & 260.3 \\
$^{88}$Y   & 1836.1 & 0.992  & 106.6 & 4.539 \\
\hline
\end{tabular}
\caption{Listing of gamma radiation emitted by the ``source cocktail'' 
and their emissivities 
at the data taking date (Nov 22, 2013). 
Gamma energies, branching fraction, and half-lives are taken from \cite{nudat}.
\added{The calculations were made assuming that the radio nuclides in the source solution
were homogeneously distributed throught the solution and that
the residue left in the syringe after the transfers was negligibly small.}
}
\label{tab:liqsrc}
\end{table}

\begin{figure}
\centering
\includegraphics[clip, trim=4cm 4cm 0cm 4cm, width=.8\textwidth]{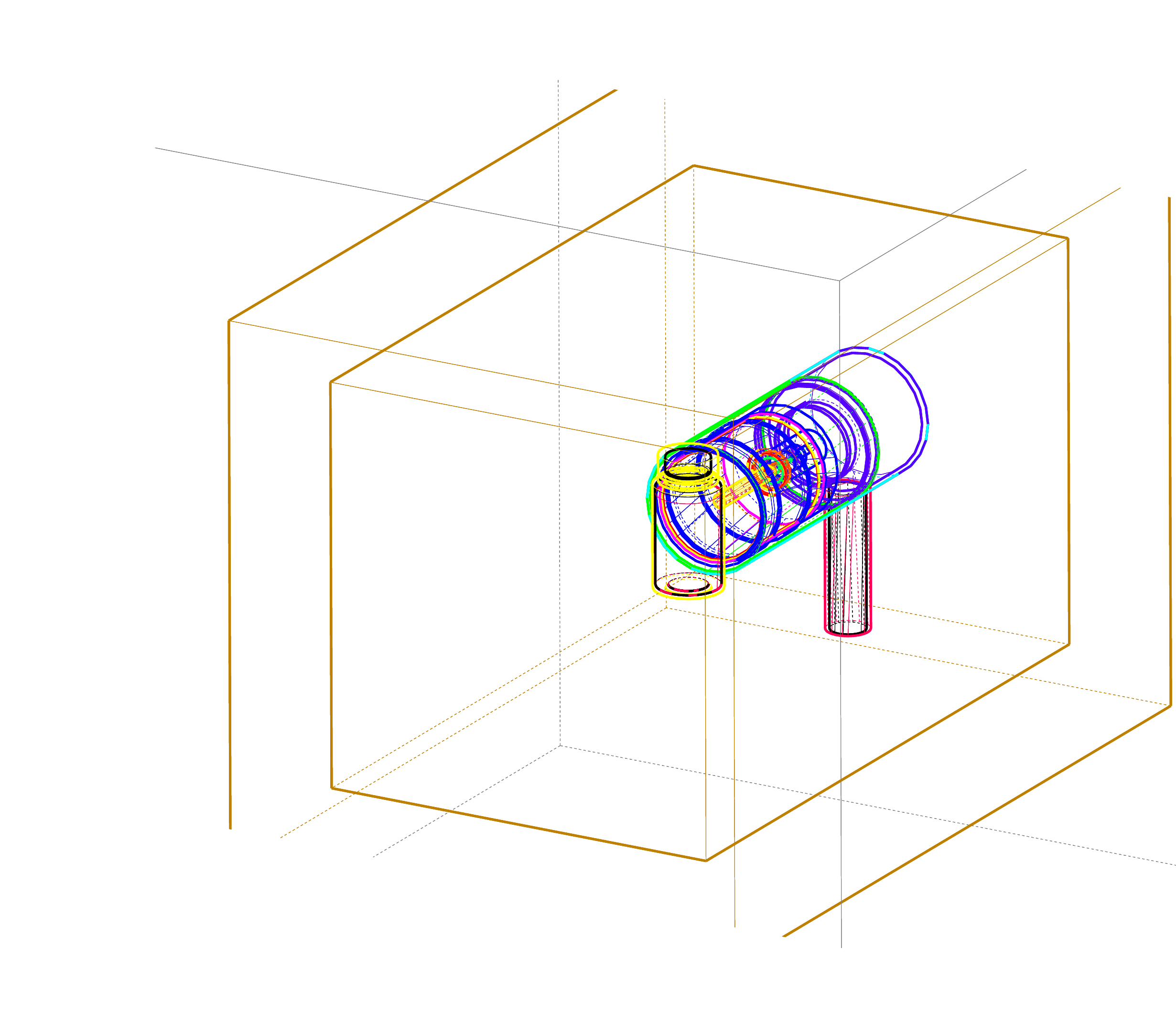} \\
\caption{Geometry for the source cocktail and the $^{210}$Pb solution.
\added{The Geant4 implementation of the PE bottle geometry was based on measurements with a caliper.}
} 
\label{fig:pebottlerendering}
\end{figure}

\begin{figure}
\centering
\includegraphics[width=\textwidth]{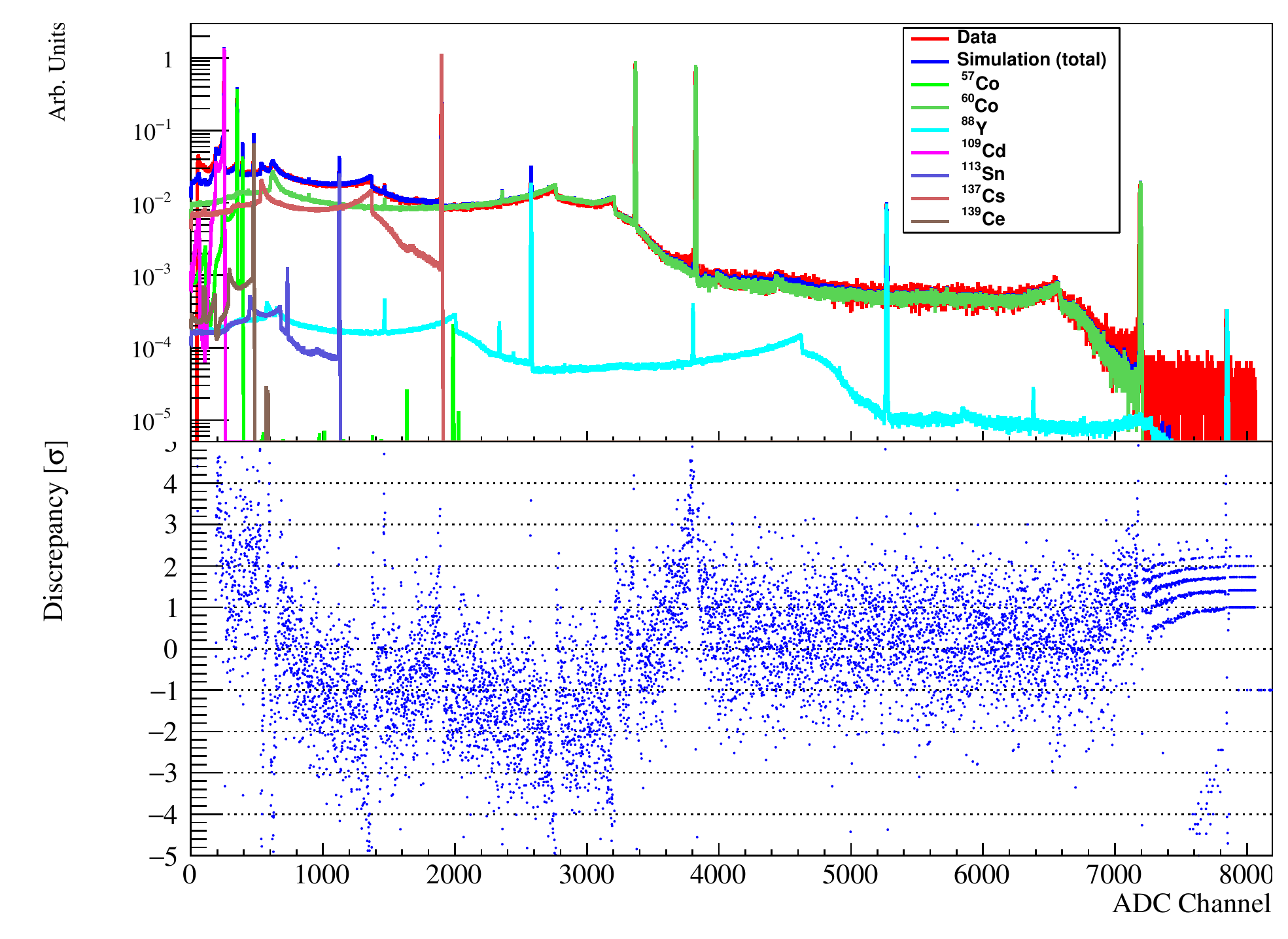} \\
\caption{The upper portion shows a comparison between (blue) simulated and (red) measured spectra for the ``source cocktail''.
The simulated spectra of the individual isotopes are in various colors indicated in the legend.
The lower portion shows the discrepancy between measurement and simulation, defined as 
$\frac{x_i-y_i}{\sqrt{\sigma_{x_i}^2+\sigma_{y_i}^2}}$ 
where 
$x_i$ and $y_i$ are the contents of the $i$th bin of the measurement and the simulation respectively, and
$\sigma_{x_i}$ and $\sigma_{y_i}$ are their respective statistical errors.
(Colors online)
} 
\label{fig:pebottlespec}
\end{figure}

\begin{figure}
\centering
\includegraphics[width=.8\textwidth]{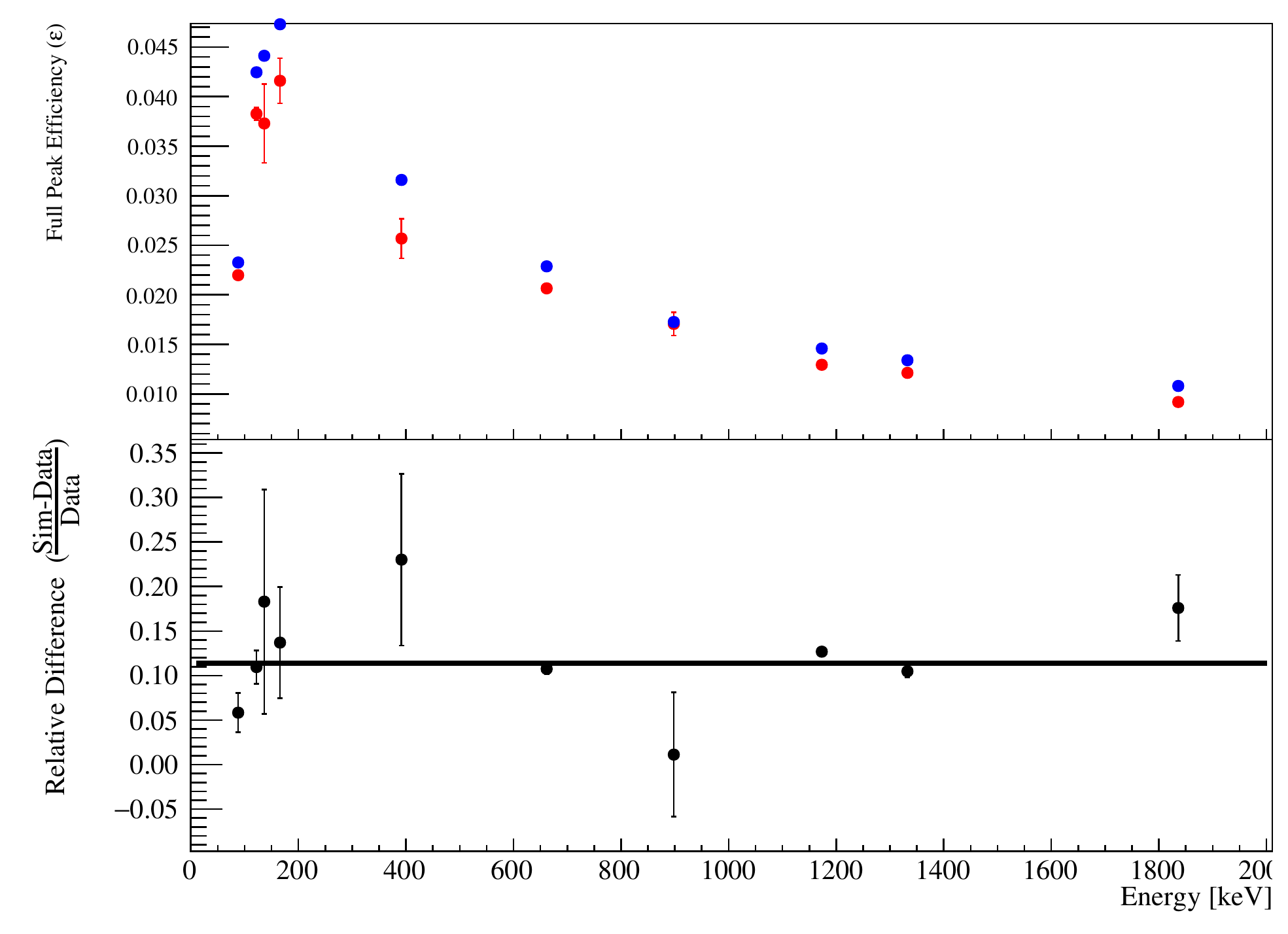}
\caption{Relative difference in full absorption peak efficiency between simulation (blue) and data (red)
for the source cocktail in GeII.
(Colors online)}
\label{fig:peaksagreement-ge2pe}
\end{figure}

\begin{figure}
\centering
\includegraphics[width=.8\textwidth]{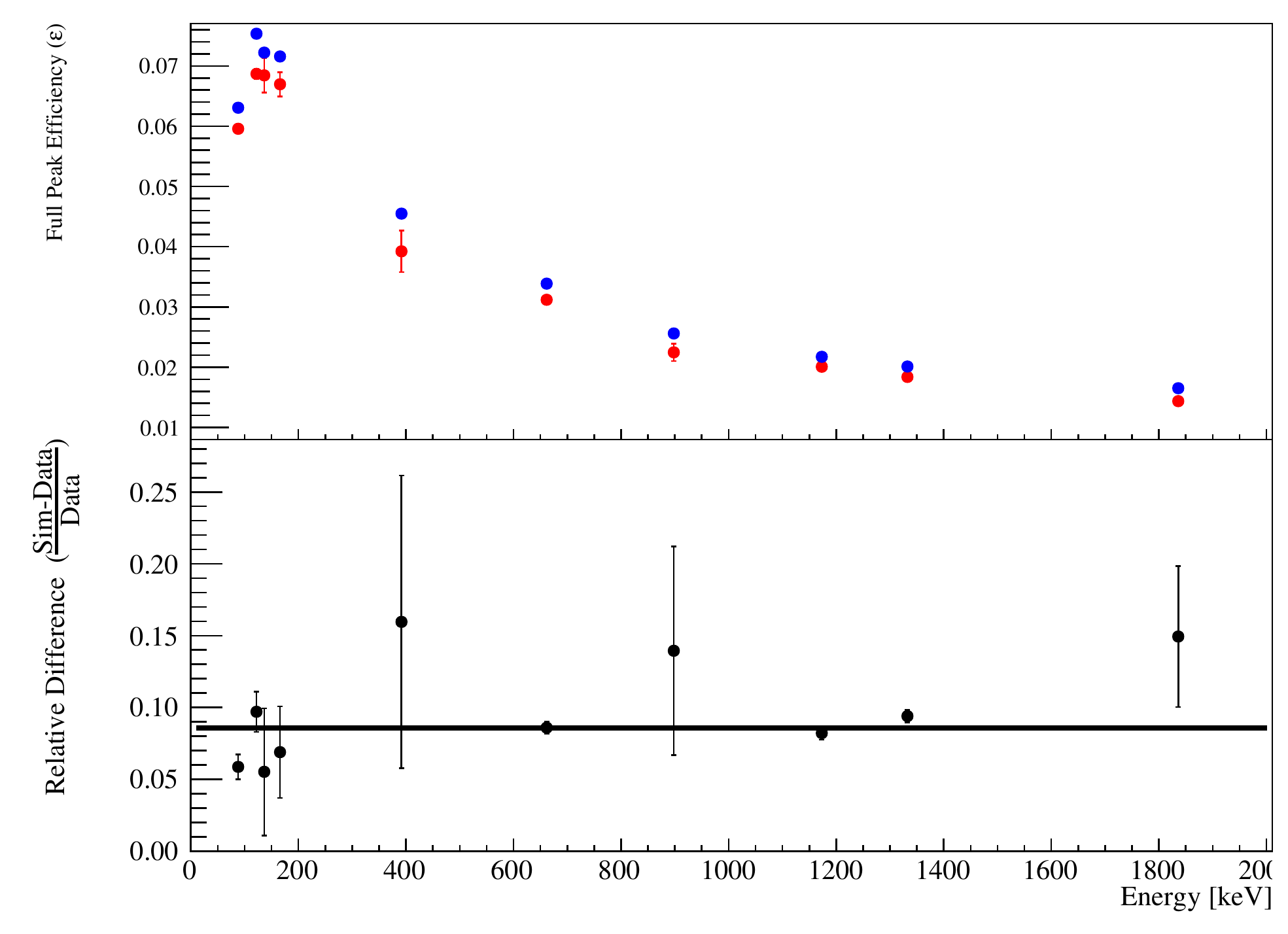}
\caption{Relative difference in full absorption peak efficiency between simulation (blue) and data (red)
for the source cocktail in GeIII.
(Colors online)}
\label{fig:peaksagreement-ge3pe}
\end{figure}

The source cocktail was counted in GeII and GeIII separately in the geometry shown in Figure \ref{fig:pebottlerendering}.
Figure \ref{fig:pebottlespec} shows a comparison between the measured and simulated spectra for GeIII.
The full absorption peak efficiencies for GeII and GeIII
are shown in Figures \ref{fig:peaksagreement-ge2pe} and \ref{fig:peaksagreement-ge3pe},
respectively. 
The differences between the measurement and simulation, 
calculated the same way as described in Section \ref{sec:posdep},
are (11.4 $\pm$ \added{0.3})\% and (8.6 $\pm$ \added{0.2})\%
for GeII and GeIII respectively (also tabulated in Table \ref{tab:syserr}.)

\subsection{Low energy region: $^{210}$Pb(NO$_3$)$_2$ solutions}

The performance of the model for GeIII at low energy ($E_\gamma = 46.54$ keV) was validated with three $^{210}$Pb sources
prepared by diluting a standard stock solution of lead-210 nitrate ($^{210}$Pb(NO$_3$)$_2$) from Eckert \& Ziegler
with BDH ARISTAR PLUS\textregistered~ nitric acid (HNO$_3$) for trace metal analysis from VWR.
This validation was not performed for GeII, 
which is equipped with a copper end cap,
as 
its detection efficiency of a photon with such a low energy 
was expected to be too low to be useful.

The sources were prepared as follows:
\begin{enumerate}
\item 54.1 mL of 70\% HNO$_3$ was diluted to 1M with 795.8 g of deionized water.
\item 5 $\mu$L of the $^{210}$Pb(NO$_3$)$_2$ solution was added to 410.4 g of the 1M HNO$_3$.
\item The diluted $^{210}$Pb(NO$_3$)$_2$ solution was filled into three 125 mL PE bottles,
the same kind as the ``source cocktail''. 
\end{enumerate}
The bottles were labeled A1, B1 and C1.
Given that the specific activity of the stock solution was \added{3.72 $\pm$ 0.15} kBq/g on April 1, 2017
\added{(derived from the calibration certificate supplied by the manufacturer)}, 
the specific activity of the \added{diluted sample solutions} was calculated to be \added{46.8 $\pm$ 2.1} Bq/kg
\added{on the same day}.

The samples were counted in GeIII in the geometry shown in Figure \ref{fig:pebottlerendering}. 
The results obtained in this test \added{(tabulated in Table \ref{tab:pbno32}) }are consistent with the expected specific activity to within 7\% \added{or about 0.7 standard deviations}.

\begin{table*}
\centering
\begin{tabular}{|c|c|c|c|c|}
\hline
Sample & Count rate   & Net sample & Measured specific\\ 
label  & [counts/day] & mass [g]   & activity [Bq/kg] \\  
\hline
A1     &  309.5 $\pm$ 14.4 & \added{133.1} & \added{48.4 $\pm$ 7.7} \\
B1     &  317.0 $\pm$ 14.6 & \added{130.3} & \added{50.7 $\pm$ 8.0} \\
C1     &  323.7 $\pm$ 13.9 & \added{130.9} & \added{51.5 $\pm$ 8.1} \\
\hline
\added{Mean} & & & \added{50.2 $\pm$ 4.6}\\
\hline
\end{tabular}
\caption{\added{Data-derived} specific activity of $^{210}$Pb in the three \added{tested sample solutions} 
using measured data and efficiency 
parametrization from simulation.
The expectation value for this \added{specific} activity is
\added{$46.8 \pm 2.1$} Bq/kg (statistical uncertainty only).
}
\label{tab:pbno32}
\end{table*}

\subsection{Cross-check among Ge analysis teams: Spiked silica samples}

This cross-check was performed as a validation among the three Ge analysis teams 
within the nEXO collaboration,
who operate the following detectors:
1) GeII and GeIII at the University of Alabama,
2) the detector at \added{Vue-des-Alpes} (VdA) operated by the University of Bern \cite{vda}, and
3) the PGT detector at SNOLAB.
In addition to the different physical locations of the Ge detectors,
there are also differences in the analysis methods among the teams as listed in Table \ref{tab:sops}.
For example, the energy scale is linear for GeIII and VdA, but quadratic at SNOLAB.
Also, the detection efficiency is modeled as Equation \ref{eq:uaeff} for GeIII, but as the following equation for SNOLAB.
\begin{equation}
\label{eq:snoeff}
\varepsilon(E) = 10^{\nu_1 + \nu_2\cdot E + \nu_3\cdot E^{-1} + \nu_4\cdot E^{-2} + \nu_5\cdot E^{-3} + \nu_6 \cdot E^{-4} },
\end{equation}
where $\nu_1$, $\nu_2$, ... ,$\nu_6$ are free varying parameters.
The VdA team does not use any analytical model, instead 
efficiencies between simulated energy points are linearly interpolated.
This exercise is to ensure that the Ge analyses done at various institutes are consistent with each other.
As such, it serves as an additional consistency test for the efficiency correction 
derived from the numerical detector model discussed in the paper.


\begin{table*}
\centering
\begin{tabular}{|L{2cm}|L{4cm}|L{4cm}|L{5cm}|}
\hline
Analysis Team & GeIII & VdA & SNOLAB \\
\hline
\multicolumn{4}{|c|}{Data taking} \\
\hline
Typical length of runs
& 24 hours 
& 24 to 72 hours
& 5 or 10 minutes
\\
\hline
Radon mitigation
& First 24 hours of data is discarded.
& First 24 hours of data is discarded.
& Sample preconditioned with Radon-poor air
\\
\hline
\multicolumn{4}{|c|}{Energy scale calibration} \\
\hline
Model
& Linear function
& Linear function
& Quadratic function
\\
\hline
\multicolumn{4}{|c|}{Continuum background subtraction} \\
\hline
Method
& Count rates are obtained by 
fitting the peak and the continuum 
at the same time
with a Gaussian + linear
model. The count rate is the integral
of the Gaussian divided by the measurement time.
& Count rates are obtained by 
integrating over a width of 2 FWHM around the peak. 
The continuum is linearly extrapolated and then 
subtracted from the peak integral.
& Count rates are obtained by  
integrating over a region of a certain width around the peak. 
The continuum is estimated by integrating 
over a region of the same width to left and another to the right 
of the peak.
The average of the two is 
subtracted from the peak integral.
\\
\hline
\multicolumn{4}{|c|}{Detection efficiency calculation} \\
\hline
Simulation tool
& GEANT4
& GEANT4
& GEANT4
\\
\hline
$\varepsilon(E)$
& Efficiencies at various energies are calculated by simulation. They are then fitted with Equation \ref{eq:uaeff}.
& Efficiencies at various energies are calculated by simulation. 
Efficiencies at intermediate energies are linearly interpolated as needed.
& Efficiencies at various energies are calculated by simulation. They are then fitted with Equation \ref{eq:snoeff}.
\\
\hline
\end{tabular}
\caption{Comparison of the standard data taking and analysis procedure among the three Ge analysis teams.}
\label{tab:sops}
\end{table*}

For this check, three samples were prepared at SNOLAB by 
adding measured amounts of three standard reference materials, purchased from the International Atomic Energy Agency (IAEA):
RGU-1 (containing uranium), 
RGTh-1 (containing thorium), and
RGK-1 (containing potassium), to silica (Fisher Scientific catalog number S-153, lot number 111211).
The actual spike amounts were blinded from the analysis teams.

The samples were prepared in the following way \cite{silica}:
\begin{enumerate}
\item Three 2 L PE bottles were filled with about 1460 grams of inactive silica at SNOLAB.
\item About 60 grams of RGU-1, RGTh-1 and RGK-1 were respectively added to individual PE bottles with silica.
\item Each of the PE bottles was mixed for 24 hours to ensure thorough mixing.
\item From each of the PE bottles, 4 grams of the mixture was transferred to a 3 mL Teflon bottle.
The three Teflon bottles were labeled U2, T2, and K2 corresponding to the mixtures containing
RGU-1, RGTh-1 and RGK-1 respectively. 
\end{enumerate}


The samples were then counted and analyzed using the GeIII, VdA, and SNOLAB detectors.
The results from the three analysis teams are consistent with each other
as can be seen in Table \ref{tab:silica}.



\begin{table*}
\centering
\begin{tabular}{|l|l|c|c|c|c|c|c|c|}
\hline
Sample     & Isotope    & GeIII        & VdA          & SNOLAB \\
\hline
U2 & $^{238}$U  & $180 \pm 2$  & $164 \pm 2$  & $180 \pm 5$ \\
T2 & $^{232}$Th & $115 \pm 2$  & $108 \pm 1$  & $116 \pm 4$ \\
K2 & $^{40}$K   & $341 \pm 10$ & $292 \pm 14$ & $377 \pm 30$ \\
\hline
\end{tabular}
\caption{Specific activities in Bq/kg of U2, T2 and K2 as measured by the three analysis teams. 
The errors are statistical only. Systematic errors are expected to be $\sim$10\% for all three teams.}
\label{tab:silica}
\end{table*}

\section{Conclusion}
\label{sec:conc}

The GEANT4 models for GeII and GeIII have been validated 
in the energy range and the geometrical configurations relevant to typical radioassay measurements. 
\added{By rounding up the largest differences 
in measured and simulated detection efficiencies for the button sources and the source cocktail
shown in Table \ref{tab:syserr},}
the systematic uncertainties of the models for GeII and GeIII are 
\added{conservatively} estimated to be $\sim$12\% and \added{$\sim$9\%} respectively.



\section*{Acknowledgements}

This research was supported in part by DOE Office of Nuclear Physics under grant number DE-FG02-01ER41166 and the NSF under award number 0923493.


RHMT was supported in part by the Nuclear-physics, Particle-physics, Astrophysics,
and Cosmology (NPAC) Initiative, a Laboratory Directed Research
and Development (LDRD) effort at Pacific Northwest National Laboratory
(PNNL). PNNL is operated by Battelle for the U.S. Department of Energy
(DOE) under Contract No. DE-AC05-76RL01830.

The authors acknowledge the support of the Natural Sciences and Engineering Research Council of Canada (NSERC).
The authors would like to thank SNOLAB for providing support in infrastructure and personnel,
and the nEXO collaboration for providing the opportunity to perform this study.



\begin{thebibliography}{00}


\bibitem{geant4}
S. Agostinelli et al.,
GEANT4 -- a simulation toolkit,
Nucl. Instr. and Methods A506, 3, 250-303 (2003)

\bibitem{knoll}
G. F. Knoll,
Radiation Detection and Measurement, 3rd Edition. 
(2000)

\bibitem{xcom}
M. J. Berger et al., 
XCOM: Photon Cross Section Database (version 1.5),
\url{http://physics.nist.gov/xcom}, accessed on October 30, 2018. 

\bibitem{nudat}
A. A. Sonzogni, 
NuDat 2.0: Nuclear Structure and Decay Data on the Internet,
AIP Conference Proceedings 769, 574 (2005).
\url{https://www.nndc.bnl.gov/nudat2}


\bibitem{vda}
P. Weber, B. A. Hofmann, T. Tolba, and J.-L. Vuilleumier,
A gamma-ray spectroscopy survey of Omani meteorites,
Meteoritics \& Planetary Science 52, Nr 6, 1017–1029 (2017)


\bibitem{silica}
B. Cleveland, Preparation and Properties of
SNOLAB Radiological Sources
SRS-12-004
SRS-12-005
SRS-12-006 --
Calibration Sources for Ge Well Detector, SNOLAB document, dated 28 March 2013.

\end{thebibliography}


\end{document}